\documentclass[prl,showpacs,twocolumn,longbibliography]{revtex4-1}

\usepackage{epsfig}
\usepackage{color}

\usepackage{graphicx}

\usepackage{amssymb,amsfonts,amsmath}
\usepackage{amssymb,amssymb,graphics,graphicx,amsfonts,amsmath,empheq}
\usepackage{pdfpages}

\usepackage{graphicx,color}

\newcommand{\cred}{\color{black}}

\def\e{\mathrm{e}}

\begin{document}

\title{Thermodynamics of Error Correction}

\author{Pablo Sartori$^1$ and Simone Pigolotti$^2$.}
\affiliation{$^1$Max Planck Institute for the Physics of Complex
  Systems. Noethnitzer Strasse 38 , 01187, Dresden,
  Germany. $^2$Dept. de Fisica i Eng. Nuclear, Universitat Politecnica
  de Catalunya Edif. GAIA, Rambla Sant Nebridi s/n, 08222 Terrassa,
  Barcelona, Spain.}

\begin{abstract}Information processing at the molecular scale is
  limited by thermal fluctuations. This can cause undesired
  consequences in copying information since thermal noise can lead to
  errors that can compromise the functionality of the copy. For
  example, a high error rate during DNA duplication can lead to cell
  death. Given the importance of accurate copying at the molecular
  scale, it is fundamental to understand its thermodynamic
  features. In this paper, we derive a universal expression for the
  copy error as a function of entropy production and {\cred work dissipated by
  the system during wrong incorporations}. Its derivation
  is based on the second law of thermodynamics, hence its validity is
  independent of the details of the molecular machinery, be it any
  polymerase or artificial copying device. Using this expression, we
  find that information can be copied in three different regimes. In
  two of them, work is dissipated to either increase or decrease the
  error. In the third regime, the protocol extracts work while
  correcting errors, reminiscent of a Maxwell demon. As a case study,
  we apply our framework to study a copy protocol assisted by kinetic
  proofreading, and show that it can operate in any of these three
  regimes. We finally show that, for any effective proofreading
  scheme, error reduction is limited by the chemical driving of the
  proofreading reaction.
    \end{abstract}


\maketitle


\section{Introduction}

Copying information is a fundamental process in the natural
world: all living systems, as well as the vast majority of manmade
digital devices, need to replicate information to function
properly. The quality of a copy relies on it being an accurate
reproduction of the original and can be quantified by the fraction
$\eta$ of wrongly copied bits that it contains. Errors can be provoked
by several hardware-specific causes, such as imperfections in the
copying machinery. At the molecular scale, perfect copying does not
exist as thermal fluctuations constitute a fundamental source of
error, regardless of the system. Since the reliability of the copying
process is ultimately limited by thermal noise, it must be understood
in terms of thermodynamics, as recognized by Von Neumann
\cite{von1956probabilistic}.

Therefore, a critical question is whether one can invoke the second
law of thermodynamics to establish a universal connection between the
error and physical quantities characterizing the copy process. This
issue should be addressed in a general framework, incorporating two
basic features of copying machineries. First, copying protocols often
involve several intermediate discriminatory steps used to regulate the
accuracy and speed of the process. This is a characteristic property
of both natural and artificial error-correcting protocols. For
example, accurate copying of DNA occurs via multistep reactions
\cite{johnson1993conformational}.  Second, due to the statistical nature of the
second law, one should consider cyclically repeated copy operations
rather than a single one \cite{Bennett:1982wx}. This cyclical operation
is also consistent with the behavior of polymerases when duplicating
long biopolymers.

\begin{figure}[!ht]
\centerline{\includegraphics[width=.48\textwidth]{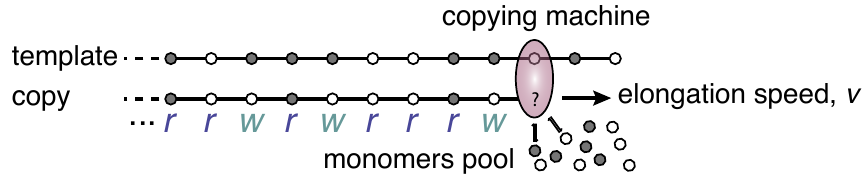}}
\caption{{\bf Template-assisted polymerization.}  The template strand
  is a pre-existing polymer made up of two different kinds of monomers
  (gray and white circles).  A molecular copying machine (red circle) assists
  the growth of a copy strand by incorporating freely diffusing
  monomers of two different types, trying to match them with those
  of the template strand. Right and wrong matches are noted $r$ and $w$.\label{fig:copol}}
\end{figure}

To understand the thermodynamics of copying, we introduce a general
framework where both the copying protocol can be arbitrarily complex
(as in models describing biochemical reactions
\cite{hopfield1974kinetic,ninio1975kinetic, murugan2012speed,murugan2014discriminatory}) and copy
operations are cyclically repeated (as in models inspired by the
physics of polymer growth
\cite{bennett1979dissipation,andrieux2008nonequilibrium,cady2009open,andrieux2009molecular,esposito2010extracting,sartori2013kinetic,andrieux2013information,gaspard2014kinetics}).
Our framework describes template-assisted growth of a copy polymer (or
``tape'', see \cite{sharma2012template}) aided by a molecular machine, see
Fig. \ref{fig:copol}.  Gray and white circles represent two different
monomer types.  The molecular machine, represented as a red circle in
the figure, is situated at the tip of the copy strand and tries to
match freely diffusing monomers with corresponding ones on the
template. When a free monomer arrives at the tip, the machine
transitions through a network of intermediate states to determine
whether to incorporate or to reject it.  Incorporation is more
likely if the matching is right, i.e. the color of the 
monomer matches that of the template, than if it is wrong. On average,
the copy strand elongates at a speed $v\ge0$ and accumulates errors
with probability $\eta$.

\begin{figure*}[hbt]
\centerline{\includegraphics[width=.96\textwidth]{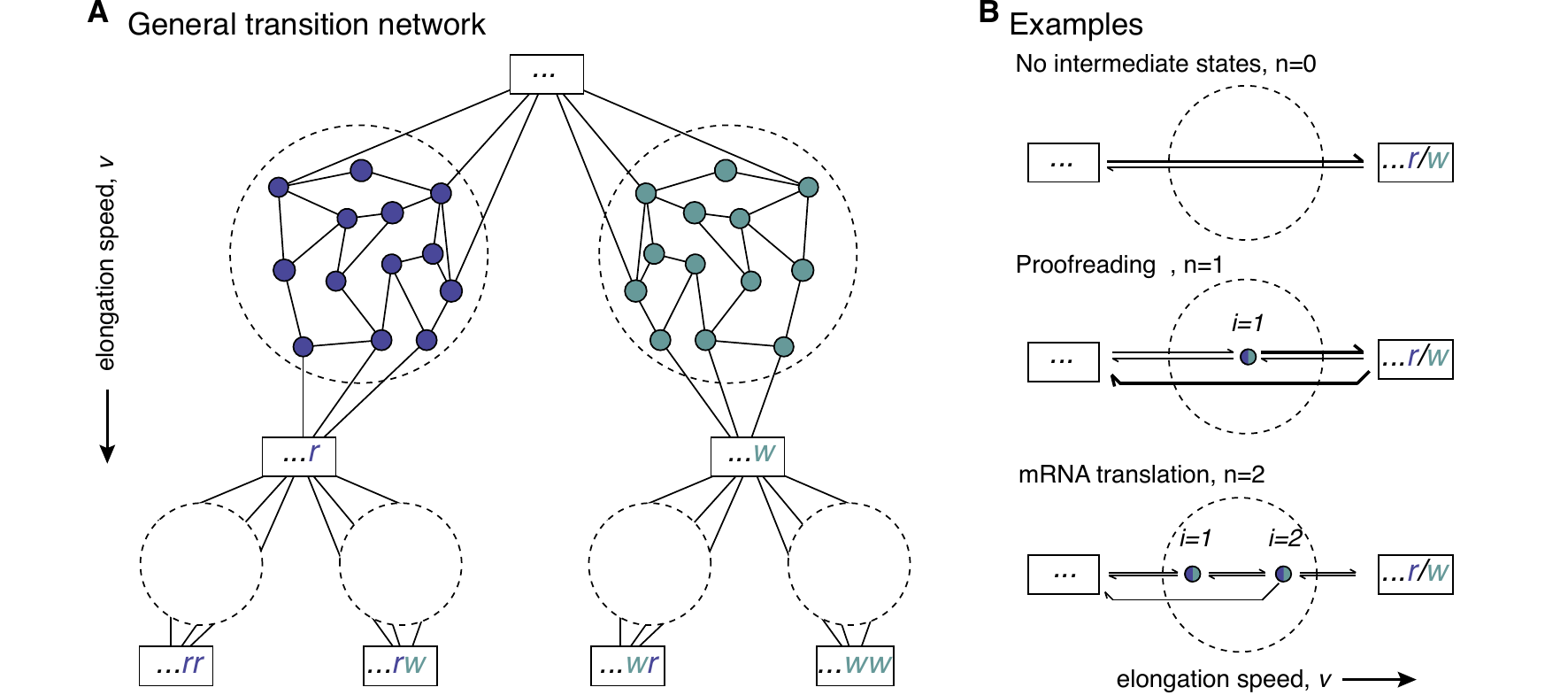}}
\caption{{\bf Transition network of template-assisted polymerization
    and examples.} {\bf A.} State space of the template-assisted
  polymerization model.  Monomer incorporation occurs via a network of
  intermediate states represented inside the dashed circles. The two
  colors distinguish networks leading to incorporation of right and
  wrong monomers. The structure is repeated in a tree-shaped structure
  as the polymer grows by addition of more and more monomers.  {\bf
    B.} Examples of networks of intermediate states. First example: 
  template-assisted polymerization without intermediate states (see
  e.g. \cite{bennett1979dissipation,sartori2013kinetic,andrieux2009molecular,esposito2010extracting,gaspard2014kinetics}).
  Second example: kinetic proofreading, where after an intermediate
  state a backwards driven pathway removes errors to improve the
  overall accuracy of the copy
  \cite{hopfield1974kinetic,sartori2013kinetic}. Third example: mRNA
  translation, where the three copying steps represent initial
  binding, GTP hydrolysis and final accommodation; a
  proofreading reaction is also present \cite{johansson2008rate}.
\label{fig:general}}
\end{figure*}

Close to thermodynamic equilibrium the process becomes  very slow, $v
\rightarrow 0$.  The error is then $\eta_{\rm
  eq}\approx\exp[-(\Delta E^{w}-\Delta E^{r})/T]$, determined by the
energy changes $\Delta E^{r}$ and $\Delta E^{w}$ of right and wrong
monomer incorporation and independently of the copying protocol.  In
this case, the error can be reduced by increasing the gap $(\Delta
E^{w}-\Delta E^{r})$, in agreement with Bennett's idea that cyclic
copying can be performed near equilibrium with arbitrary precision
\cite{Bennett:1982wx,sartori2013kinetic}. This mechanism is however
unpractical, for example due to the low speed limitation. Instead,
typical molecular machines spend chemical energy to copy at a finite
speed and out of thermodynamic equilibrium. Non-equilibrium copying
protocols can also reduce the error far below its equilibrium
value. For example, the equilibrium estimate for the error in DNA
duplication is $\eta_{\rm eq}\sim10^{-2}$, where the actual observed
error is $\eta\sim10^{-9}$ \cite{johnson1993conformational}.  An important
non-equilibrium mechanism underlying error correction is kinetic
proofreading, which feeds on chemical energy to preferentially undo
wrong copies \cite{hopfield1974kinetic,ninio1975kinetic,bennett1979dissipation}. Other
non-equilibrium mechanisms such as induced fit \cite{pape1999induced} and
kinetic discrimination \cite{cady2009open,sartori2013kinetic} complement
kinetic proofreading to underpin the high accuracy of replication in
biological systems.

In this work we demonstrate that, for the broad class of processes
depicted in Fig.~\ref{fig:copol}, a direct relation links copy errors
with non-equilibrium thermodynamic observables {\cred characterizing
  incorporation of errors}. In particular, at
fixed work budget, the error decreases exponentially with the total
entropy produced per wrongly copied bit.  This relation is completely
general, in contrast with conditions setting hardware-specific minimum
errors $\eta_{\rm min}$ that characterize each particular copying
protocol.  When studying wrong matches alone, three copying regimes
can be identified: {\it error amplification}, where energy is invested
in increasing the error rate; {\it error correction}, where energy is
invested in decreasing the error rate; and {\it Maxwell demon}, where
the information contained in the errors is converted into work.  We
conclude by studying the specific copying protocol of kinetic
proofreading.  We show that proofreading can operate in all these
three regimes. Furthermore, for a broad class of proofreading
protocols, we show that error reduction is limited by the chemical
energy spent in the proofreading reaction.

\section{Results}
\subsection{Template-assisted polymerization}

We start our discussion by detailing the stochastic dynamics of the
template-assisted polymerization process sketched in Fig.
\ref{fig:copol}.  Its transition network is represented in
Fig. \ref{fig:general}A. The rectangles correspond to the states of the
system after the copying machine finalized incorporation of a
monomer. We denote them with a string such as $\dots rrwr$, which
refers to a particular sequence of right and wrong matches (see also
Fig. \ref{fig:copol}). Dashed circles encloses sub-networks of $n$
intermediate states, characteristic of the copying protocol.  The intermediate
states, represented as blue/green circles for right/wrong matches in
Fig. \ref{fig:general}A, are used by the machine to process a tentatively
matched monomer and decide whether to incorporate it or not.  We note
intermediate states as  $\dots rrwr r_i$, with $1\le
i\le n$, and analogously for wrong monomers. A copying protocol is
fully specified by the topology of the sub-networks, assumed to be the
same for right and wrong matches, and the kinetic rates $k_{ij}^r$
for right matches and $k_{ij}^w$ for wrong ones. Differences in
the rates are responsible for discrimination. Possible examples of
sub-networks of increasing complexity are represented in Fig.  \ref{fig:general}B.

Because of thermal fluctuations induced by the environment at
temperature $T$, all kinetic transitions are stochastic. The
states are thus characterized by time-dependent probabilities $P(\dots
r)$, $P(\dots w)$, $P(\dots r_i)$ and $P(\dots w_i)$. Their evolution
is governed by a set of master equations which can be solved at steady
state, see {\em Methods}. Key to the solution is to postulate that
errors are uncorrelated along the chain, so that
$P(\dots)\propto\eta^{N^w}(1-\eta)^{N-N^w}$, where $N$ is the length
of the chain and $N^w$ is the total number of incorporated wrong
matches. The error $\eta$ can then be determined via the
condition
\begin{equation}
\label{eq:error}
\frac{\eta}{1-\eta}=\frac{v^w(\eta)}{v^r(\eta)},
\end{equation}
where $v^r$ and $v^w$ are the average incorporation speeds of right and wrong monomers, respectively. They represent the average net rates at which right and wrong monomers are incorporated in the copy. The net elongation speed $v$ is the sum of these two contributions, $v=v^r+v^w$. Substituting the solution for $P(\ldots)$ 
into the master equations leads to explicit expressions for $v^w$ and $v^r$ as a
function of the error and all the kinetic rates. In this way,
Eq. (\ref{eq:error}) becomes a closed equation for the only
unknown $\eta$.  Note that Eq.~\ref{eq:error} and the definition of $v$ imply $v^r=(1-\eta)v$ and $v^w=\eta v$.


\subsection{Thermodynamics of copying with errors}

\begin{figure}[!ht]
    \centerline{\includegraphics[width=.48\textwidth]{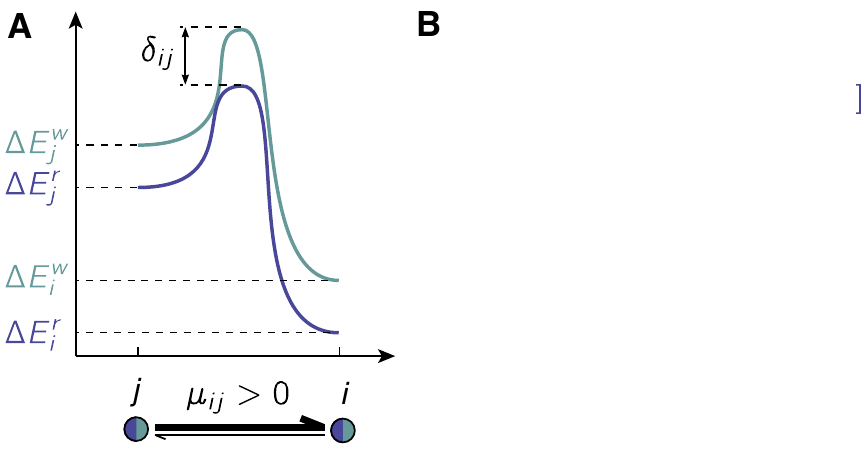}}
    \caption{{\bf Energy landscape and kinetic rates.} {\bf
        A} Energetic diagram of a single transition in the reaction
      network. {\bf B} Corresponding kinetic rates. The transition
      $j\to i$ can be driven by energy differences and the chemical
      driving $\mu_{ij}$.  Transitions involving a right and a wrong
      monomer can be characterized by different kinetic barriers
      $\delta_{ij}$, as well as different energetic landscapes $\Delta
      E_j^w \neq \Delta E_j^r$. The bare rate $\omega_{ij}$ is the
      inverse characteristic time scale of each reaction.
              \label{fig:rates}}
\end{figure}

The kinetic rates $k_{ij}^r$ and $k_{ij}^w$ are determined by the
energy landscape of the system, the chemical drivings $\mu_{ij}$ of
the reactions, and the temperature $T$ of the thermal bath, as
represented in Fig. \ref{fig:rates}A.  The chemical drivings
  represent the difference in chemical potential of reactions, such as
  ATP hydrolysis, fueling the transitions $j\to i$. The energy
differences of an intermediate state respect to the state before the
candidate monomer incorporation are $\Delta E_i^r=E(\dots
r_i)-E(\dots)$, and similarly for wrong incorporation; the energy
changes after {\em finalizing} incorporation of a
monomer are  $\Delta E^r=E(\dots r)-E(\dots)$ and analogously
for wrong matches.   Note
  that these {\it energies} are in a strict sense free energies as
  they might depend, for example, on the monomer concentrations in the
  cell. Energetic discrimination can be exploited when the wrong
match is energetically more unstable than the right one, $\Delta
E^w\ge \Delta E^r$.  In addition, wrong matches can also be
discriminated kinetically, i.e. by exploiting different activation
barriers $\delta_{ij}$ in the transitions performed by the machine when
a right monomer is bound.  In general, complex copying protocols can
combine both these mechanisms
\cite{sartori2013kinetic,zaher2009fidelity}.  Full expressions of the
rates are summarized in Fig. \ref{fig:rates}B.

Given a steady-state elongation speed $v$, the chemical drivings
perform an average work per added monomer $\Delta W = \sum_{\langle
  ij\rangle}\mu_{ij}(J^r_{ij}+J^w_{ij})/v$, where $J_{ij}^r$ and
$J_{ij}^r$ are probability fluxes (see also {\em Methods}).  Further,
{\cred the free-energy change per added monomer at equilibrium
would be} $\Delta F_{\rm eq} = - T\log (e^{-\Delta E^r/T}+e^{-\Delta
  E^w/T} )$. In the  limit $v\rightarrow 0$, the system
approaches equilibrium and the population of all states is determined
by detailed balance. This implies that the equilibrium error is
$\eta_{\rm eq} =\exp{[(-\Delta E^w+\Delta F_{\rm eq})/T]}$. When driving
the dynamics out of equilibrium, the error will in general depart from
its equilibrium value, leading to a positive total entropy
production. In {\em Methods}, we derive that the total entropy
production per copied monomer and the error are linked by the relation
\begin{equation}
T\Delta S_{\rm tot}=\Delta W - \Delta F_{\rm eq}- TD(\eta||\eta_{\rm eq})\ge0\quad ,
\label{eq:entropyprod}
\end{equation}
where $D(\eta||\eta_{\rm eq})=\eta \log(\eta/\eta_{\rm
  eq})+(1-\eta)\log[(1-\eta)/(1-\eta_{\rm eq})]$ is the
Kullback-Leibler distance between the equilibrium and non-equilibrium
error distribution, which is always non-negative and vanishes only for
$\eta=\eta_{\rm eq}$. Eq. \ref{eq:entropyprod} states that the
  average performed work is greater than the equilibrium free energy increase
  by a configurational bound, $\Delta W-\Delta F_{\rm eq} \ge
  T~D(\eta||\eta_{\rm eq})\ge 0$.  In this view, the Kullback-Leibler
  term in Eq. \ref{eq:entropyprod} can be interpreted as the
  additional free energy stored in a copy characterized by an error
  different from its equilibrium value.  This additional free energy
  can be recovered by a spontaneous depolymerization process that will
  stop once the system reaches its equilibrium error
  \cite{bennett1979dissipation}.

Eq. (\ref{eq:entropyprod}) relates the information content of the copy
with thermodynamics. However, in many relevant cases, the entropy
production is dominated  by the {\cred ``excess work'' } $\Delta W-\Delta F_{\rm eq}$, so that in practice
Eq. (\ref{eq:entropyprod}) reduces to the traditional form of the
second law. Consider for example a case in which error correction is
very effective, $\eta\ll\eta_{\rm eq}$. In this limit, the
Kullback-Leibler term tends to a constant, $D(\eta||\eta_{\rm
  eq})\to-\log(1-\eta_{\rm eq})>0$. Since usually the equilibrium
error is already small, this constant is also small,
$D(\eta||\eta_{\rm eq})\approx\eta_{\rm eq}\ll1$. The reason is that,
since errors are typically rare, their overall contribution will be
small.

To better understand the link between errors and thermodynamics,
{\cred we consider the average entropy production associated with an
  error incorporation, $\Delta S_{\rm tot}^w=\dot{S}_{\rm tot}^w/v^w$,
    where $\dot{S}_{\rm tot}^w$ is the entropy production rate coming from
      incorporation of wrong monomers only. The quantity $\Delta
      S_{\rm tot}^w$ also obeys a second-law-like inequality}
\begin{equation}
\label{eq:dsw}
T\Delta S_{\rm tot}^w =\Delta W^w-\Delta F_{\rm eq} -  T
\log(\eta/\eta_{\rm eq}) \ge 0,
\end{equation}
where $\Delta W^w=\sum_{\langle ij\rangle}J^w_{ij}\mu_{ij}/v^w$ is the average
work performed per wrong match (see {\em Methods}). Rearranging terms in
Eq. (\ref{eq:dsw}) yields a general expression for the error in
terms of thermodynamic observables
\begin{equation}
\label{eq:ESA}
\eta = \eta_{\rm eq}\exp\left[-\Delta S_{\rm tot}^w +(\Delta {W}^w-\Delta F_{\rm eq})/T\right].
\end{equation}
This result does not depend on microscopic details of the copying
protocol, such as the discrimination barriers $\delta_{ij}$.  Eq.
(\ref{eq:ESA}) provides a direct link between thermodynamic
irreversibility and accuracy of copying. It states that, given a fixed
work budget, reduction of the error beyond its equilibrium value
always comes at a cost in terms of entropy production. However, the
dependence of the error on the thermodynamic quantities is
non-trivial to derive from Eq. (\ref{eq:ESA}), as varying the work
also affects the entropy production.

\begin{figure}[!htb]
  \centerline{\includegraphics[width=.48\textwidth]{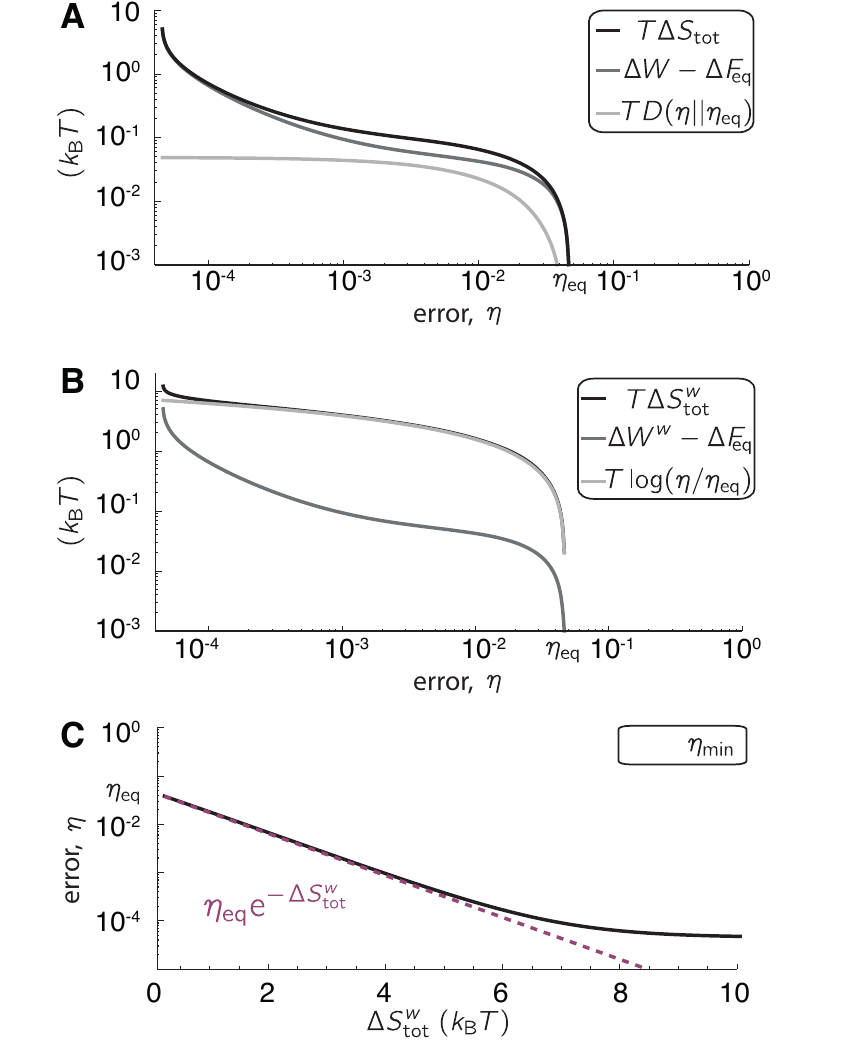}}
  \caption{{\bf Template-assisted polymerization without intermediate
      states.} {\bf A} {\cred Excess work} $\Delta W-\Delta F_{\rm eq}$, entropy
    production and Kullback-Leibler term of Eq.
    (\ref{eq:entropyprod}) as a function of the error. Notice that the
    excess work dominates over the information term. {\bf B}
    Same terms as in {\bf A}, but for wrong monomers only. In this
    case, the information term dominates the entropy production. {\bf
      C} Relation between error and entropy production of wrong
    monomers, together with thermodynamic (red, dashed) and
    hardware-specific (black, dashed) bounds.  In all panels, the
    driving $\mu_{10}$ is varied to vary the error.  Parameters are
    $\delta_{10}=10T$, $\Delta E^r_1=0$, $\Delta
    E^w_1=3T$.\label{fig:thermo}}
\end{figure}

The inequality of Eq. (\ref{eq:dsw}) reveals the existence of
three possible copying regimes: 
\begin{enumerate}
\item {\it Error amplification}, $\Delta W^w-\Delta F_{\rm eq}>0$ and
  $\eta>\eta_{\rm eq}$. In this regime, {\cred a positive excess work for
  wrong matches} leads to an error higher than its equilibrium
  value. While, in this case. dissipating energy is counterproductive in
  terms of the achieved error, it can be justified by the need of
  achieving a high copying speed.
\item {\it Maxwell demon}, $\Delta W^w-\Delta F_{\rm eq}<0$ and
  $\eta<\eta_{\rm eq}$. In this regime, the machine {\em extracts}
  work while lowering the information entropy of the chain with
  respect to its equilibrium value, $-\eta\log(\eta)<-\eta_{\rm
    eq}\log(\eta_{\rm eq})$. This regime is reminiscent of a
  Maxwell demon, since an apparent violation of the 
    second-law-like inequality, Eq. \ref{eq:dsw}, occurs from
  neglecting entropy production associated with information
  manipulation (see e.g. \cite{jarz}). Note, however, that the
  {\cred excess work associated to right matches compensates this
    term}, so that growth of a copolymer can not result in $\Delta
    W-\Delta F_{\rm eq}<0$, see Eq.~\ref{eq:entropyprod}.
\item {\it Error correction}, $\Delta W^w-\Delta F_{\rm eq}>0$ and
  $\eta<\eta_{\rm eq}$. This is an error-correction scenario in which
  work is dissipated to achieve an error lower than the equilibrium
  error. In this case, which is the most common for biological
  machines, Eq. (\ref{eq:ESA}) implies a simple bound on the
  error, $\eta \ge \eta_{\rm eq}\exp(-\Delta S_{\rm tot}^w )$.
\end{enumerate}

Given the copying protocol and the kinetic rates, the copying
machinery will achieve a certain error $\eta$ and   
operate in one of these three regimes. Varying the kinetic rates
affects both the error and the thermodynamic observables, possibly
shifting the operating regime of the machine. To better scrutinize
these aspects, we now move to considering specific protocols.

In the simplest possible example, incorporation occurs in a single
step, as sketched on the top panel of Fig.~\ref{fig:general}B (see
also
\cite{bennett1979dissipation,sartori2013kinetic,andrieux2009molecular,esposito2010extracting,gaspard2014kinetics}). It
can be shown that this protocol is always dissipative, $\Delta
W^{w}-\Delta F_{\rm eq}\ge0$. In general, wrong monomers can be discriminated
by a kinetic barrier $\delta_{10}$ and an energy difference $\Delta
E^w-\Delta E^r$ \cite{sartori2013kinetic}. If the kinetic barrier is
larger than the energy difference, $\delta_{10}>\Delta E^w-\Delta
E^r$, it can be shown that $\eta<\eta_{\rm eq}$, corresponding to {\it
  error correction}. If it is lower, then $\eta>\eta_{\rm eq}$, which
corresponds to {\it error amplification} \cite{sartori2013kinetic}.
In Fig.~\ref{fig:thermo}A we plot the different terms of the total
entropy production, Eq. (\ref{eq:entropyprod}), for the error
correction case. As discussed before, the information contribution to
the total entropy production is negligible for small errors. Instead,
note in  Fig.~\ref{fig:thermo}B that the information term of
Eq. (\ref{eq:dsw}) dominates over the work performed per wrong
matches. This implies that the universal expression for the error,
Eq.~(\ref{eq:ESA}), is very well approximated by the lower bound of
{\it error correction}, as shown in Fig.~\ref{fig:thermo}C. The error
departs from this bound only when it approaches its hardware-specific
minimum $\eta_{\rm min}\approx \e^{-\delta_{10}/T}$. Note that
  increasing $\delta_{10}$ decreases both $\eta_{\rm min}$ and the
  dissipative cost $\Delta S_{\rm tot}^{w}$ of copying at an error
  rate $\eta>\eta_{\rm min}$.

\subsection{Energetic bound to proofreading accuracy} 
\begin{figure}[!htb]
\begin{center}
        \centerline{\includegraphics[width=.48\textwidth]{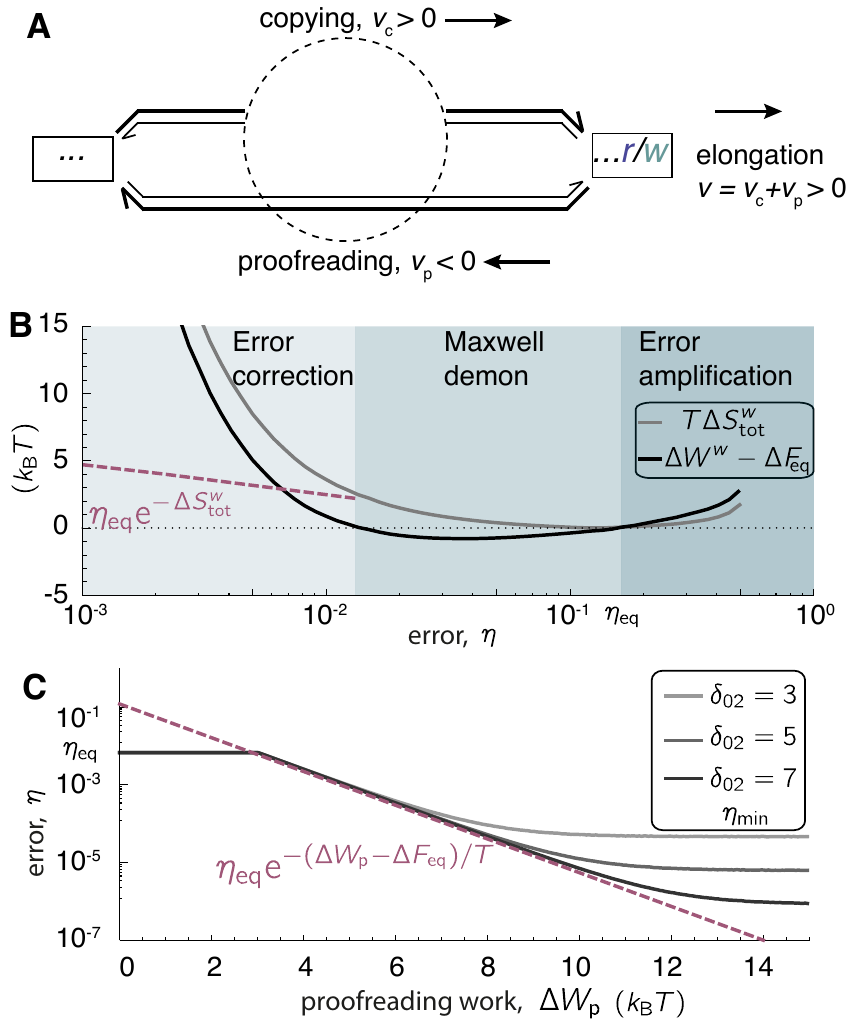}}
        \caption{{\bf Regimes and bounds of kinetic proofreading.} {\bf A} Scheme of a
          generic proofreading scheme. Copying occurs at a net
          speed $v_{\rm c}>0$ through an arbitrary reaction network
          of intermediate states. After the copy
          is finalized, a proofreading reaction removes errors at a
          speed $v_{\rm p}<0$. The net average speed is
           $v=v_{\rm c}+v_{\rm p}\ge0$  {\bf  B.} Thermodynamic
          regimes of kinetic proofreading. The model combines a copying
          scheme with one intermediate state with kinetic proofreading, as
          represented in Fig. (\ref{fig:general}B). The shaded regions
          denote the three thermodynamic regimes discussed in the
          previous section. Parameters are
          $\delta_{10}=5T$, $\delta_{21}=0$, $\delta_{02}=5T$, $\Delta
          E_2^w=\Delta E_1^w=2T$, $\Delta E_2^r=\Delta
          E_1^r=0$. {\cred We remind that states $0$, $1$,  and $2$
            represent the state before monomer incorporation, the
            intermediate state, and the final state where monomer has
            been incorporated, respectively
            (see also Fig. \ref{fig:rates}B)}. For
          each value of the error $\eta$, the other free parameters
          ($\mu_{10}$, $\mu_{21}$, $\omega_{21}$, $\omega_{02}$) are
          determined by minimizing the entropy production per copied
          wrong monomer $\Delta S_{\rm tot}^w$. 
 {\bf  C} Minimum error as a function of the proofreading work
 $\Delta W_{\rm p}=\mu_{02}$. For each curve, energies and activation barriers are
 fixed parameters as in the previous panel (except for $\delta_{02}$
 which varies, as in the captions).  For each value of $\mu_{02}$, the
 other free parameters ($\mu_{10}$, $\mu_{21}$, $\omega_{21}$,
 $\omega_{02}$) are determined by numerically minimizing the error
 $\eta$. Red-dashed and black-dashed lines represent thermodynamic and
 hardware-specific bounds, respectively.
 \label{fig:kp} }
\end{center}
\end{figure}

In kinetic
proofreading, a copying pathway that incorporates monomers at a speed
$v_{\rm c}\ge0$ is assisted by a parallel pathway which preferentially
removes wrong matches at a speed $v_{\rm p}\le0$, see
Fig.~(\ref{fig:kp}A). {\cred Hereafter the sub-index ``p'' indicates that quantities are computed only for the proofreading reaction}. To maintain a negative speed, the proofreading
reaction must be driven backward either by performing a work 
{\cred per added monomer} $\Delta W_{\rm p}$ , or by exploiting a high free energy difference $\Delta F_{\rm eq}$
between the final and the initial state. By means of proofreading, one
can achieve lower errors than those of the copying pathway alone, at
the cost of spending additional chemical driving and reducing the net
copying speed $v=v_{\rm c}+v_{\rm p}$.

We consider a proofreading protocol consisting of a copying pathway
with one intermediate step in addition to the proofreading reaction,
see middle panel in Fig.~\ref{fig:general}B.  By tuning the rates,
this model can operate in all three regimes described in the previous
section, as shown in Fig. \ref{fig:kp}B.  In particular, in the {\it
  Maxwell demon} regime, the error can be reduced up to one order of
magnitude below its equilibrium value while at the same time
extracting work from the wrong copying reaction. Very small errors are
achieved in a strongly driven {\it error correction} regime, where the
error rate satisfies $\eta \ge
\eta_{\rm eq}\exp(-\Delta S_{\rm tot}^w )$. However, at variance with
the example of the previous section, here the entropy production
becomes quickly much larger than this bound. The reason is that
  effective proofreading requires a cycle in the reaction pathway
  which fundamentally involves dissipation of
work. This dissipation, rather than the information term, dominates
the entropy production of wrong matches at low errors. This is at
variance with the single-step model of the previous section, where no
cycles are present and the configurational entropy dominates over dissipation.

To derive a better estimate of the error in proofreading, we now focus on
the rate of entropy production  {\cred during the proofreading of wrong matches} $T\dot{S}^w_{\rm p,
  tot}=-v_{\rm p}^w\Delta W_{\rm p}^w-v_{\rm p}^w[\Delta F_{\rm eq}+
T\log\left({\eta}/{\eta_{\rm eq}}\right)]$.  Using that in
proofreading $v_{\rm p}^w<0$ while $\dot{S}^w_{\rm p, tot}\ge 0$, we can
derivethe following bound for the
error
\begin{equation}
\label{eq:kpbound}
\eta\ge \eta_{\rm eq}\exp\left( -\frac{\Delta W_{\rm p}+\Delta F_{\rm eq}}{T}\right)\quad ,
\end{equation}
{\cred where we have further used that $\Delta W_{\rm p}=\Delta W_{\rm
    p}^w$  (see {\em Methods} for details) }. This equation is one of the main results of this paper. It shows that error reduction in proofreading is limited by
its energetic cost, either in the form of chemical work in the
proofreading pathway \cite{zaher2009fidelity} or free energy of the final state,
which involves performing work in the copying pathway
\cite{hopfield1974kinetic}.  Similarly to Eq. (\ref{eq:ESA}), this bound
does not depend on details of the copying protocol.  In
Fig. \ref{fig:kp}C, we show the error of the specific proofreading
model of Fig. \ref{fig:kp}B as a function of the 
proofreading work.  One can appreciate that the bound from
Eq.~\ref{eq:kpbound} is tightly met for a wide range of errors. For
very small values of $\Delta W_{\rm p}$, when $v_{\rm p}>0$ and no
proofreading occurs, the bound is not satisfied. Finally, for very
large work values, the error approaches the hardware-specific minimum
$\eta_{\rm min}$.

In this case, the value of $\eta_{\rm min}$ can be obtained from the
explicit solution of the model (see derivation in {\em Methods}).  In
the strongly driven regime, the error $\eta$ decreases at increasing
proofreading work $\Delta W_{\rm p}$. At the same time, $v_{\rm
  p}$ becomes more negative as more copies are proofread.  The minimum
error is thus obtained in the limit of vanishing elongation speed,
when the proofreading speed is negative enough to arrest copying,
$v_{\rm p}=-v_{\rm c}$.  Imposing this condition gives the
hardware-specific minimum error
\begin{equation}
\eta_{\rm min}\approx e^{(-\delta_{10}+\delta_{02}-\Delta  E^w +\Delta
  E^r)/T}\quad .
\label{eq:hardware}
\end{equation}
This expression shows that the error of the first copying step,
approximatively equal to $e^{-\delta_{10}/T}$ because of the large
kinetic barrier, is reduced by a factor $e^{(\delta_{02}-\Delta
  E^w+\Delta E^r)/T}$ due to the additional discrimination of the
proofreading reaction.

\section{Discussion}

In this paper, we analyzed template-assisted polymerization, where
copies are cyclically produced by an arbitrary complex reaction
network. This broadly extends {\cred Bennett's original
  copolymerization model \cite{bennett1979dissipation} and further
  studies
  \cite{andrieux2008nonequilibrium,cady2009open,andrieux2009molecular,esposito2010extracting,sartori2013kinetic,andrieux2013information,gaspard2014kinetics})}
where monomer incorporation occurs in a single step. In particular,
the results presented here allow for analyzing the thermodynamics of
realistic biological copying protocols, where a complex reaction network is
responsible for error correction.

At variance with models for the copy of a single monomer
\cite{hopfield1974kinetic,ninio1975kinetic,
  murugan2012speed,murugan2014discriminatory}, in template-assisted
polymerization {\cred the number of possible states of the chain grows
  exponentially at steady-state.  This exponential increase} causes
the appearance of an information term in the formula for the total
entropy production, Eq. \ref{eq:entropyprod}.  A similar term appears
in the context of Landauer principle out of equilibrium
\cite{esposito_landauer}, and was interpreted as the amount of
information necessary to shift from the equilibrium distribution to
the non-equilibrium one. Eq.  \ref{eq:entropyprod} should not be
confused with a formally similar one {\cred derived by Gaspard and
  Andrieux} \cite{andrieux2008nonequilibrium}, whic represents a
physically different quantity, i.e. the entropy of the copy given the
template. {\cred This difference is physically important: the
  information term in Eq. \ref{eq:entropyprod} can be thought of as a
  measure of distance from equilibrium, as it is equal to zero at
  equilibrium and positive otherwise. In contrast, the information
  term in Gaspard and Andrieux's formula goes to zero only in the
  limit of vanishing error rate.}

The main result of this paper is that, thanks to the explicit
dependence on the error, the second law of thermodynamics can be used
to obtain general expressions and bounds on the copy error. This
allows us to identify three different copying regimes: error
amplification, error correction, and Maxwell demon, all of which can
be achieved by kinetic proofreading.

Considering cyclic copying is analogous to considering cyclic
transformation when studying the efficiency of thermodynamic
engines. Besides being the natural choice to properly describe the
thermodynamics of the process, template-assisted polymerization 
allows for out-of-equilibrium copying regimes which are
absent in single-monomer models.  For example, a lower bound to the
error analogous to Eq. \ref{eq:kpbound} is generally valid in
closed networks \cite{ehrenberg_proof,qian_noise}. In
template-assisted polymerization, this limit can be broken when the
proofreading reaction reverts its flux, as seen in Fig. \ref{fig:kp}D
for small values of the work.

We briefly discuss the relevance of our results for interpreting
experimental data.  Many biological copying pathways are driven by the
hydrolysis of one single GTP molecule. The chemical work spent in this
process is $\Delta\mu=\Delta\mu^{0}+k_{\rm
  B}T\log\left(\frac{[GTP]}{[GDP][Pi]}\right)$. Taking as reference
the bare potential of ATP, $\Delta \mu^{0}=14.5k_{\rm B }T$, and
typical concentrations $[GTP]=1$mM, $[GDP]=0.01$mM and $[Pi]=1$mM, we
obtain $\Delta \mu_{GTP}\approx 20k_{\rm B}T$. In a protocol involving
proofreading, this information and Eq.~\ref{eq:kpbound} can be used to
set a lower bound for the error.  Assuming that the energy of GTP is
all spent to increase the free energy of the chain, $\Delta
F\approx\Delta\mu_{GTP}$, we obtain that the total error reduction is
$\eta/\eta_{\rm eq}\ge10^{-9}$.  The value of this bound is
smaller than typically observed errors, which reasonably suggests
that not $100\%$ of the energy of hydrolysis is utilized to
increase the free energy of the system.

Given the flexibility of our framework, many complex copying
mechanisms studied in the literature as non-cyclic processes
\cite{johansson2008rate,pape1999induced,zaher2009fidelity} can be directly considered as
template assisted polymerization problems and studied from the point of view of
thermodynamic efficiency.  One limitation of our treatment is the lack
of long-term memory: while processing a monomer, the machine does 
not keep track of the past errors encountered along the chain. A more
general scheme could exploit correlations in the template sequence to
reduce the error. An example of this is backtracking \cite{galburt2007backtracking,depken,mellenius2015dna},
where regions of the template containing many errors are
entirely reprocessed. Generalization of template-assisted
polymerization to these cases will be the subject of a future study.

The thermodynamic relations derived in this paper fundamentally limit
the capabilities of stochastic machines to reduce and proofread
errors, and are reminiscent of similar bounds derived for adaptation
error in sensory systems \cite{lan2012energy} {\cred It will be of
  interest to understand whether our results can be applied to error
  correction in sensing.  For example, it is known that sensory
  pathways exploit proofreading both in chemosensing by isolated
  receptors \cite{morathierry} or cooperative ones
  \cite{lalanne2015chemodetection}}.  Clarifying the links between
these problems will constitute an important step towards formulating
general thermodynamic principles \cite{Parrondo2015} limiting the
accuracy of non-equilibrium information-processing.

\section{Methods}
\subsection{Steady-state solution of template-assisted polymerization} In this section, we briefly outline how to solve the
template-assisted polymerization model. We start by writing the master
equations governing the evolution of probabilities of all main states
$P(\dots)$, and those of the intermediate states $P(\dots r_i)$ and
$P(\dots w_i)$. The probability flux between two arbitrary intermediate states $\dots r_j$ and
$\dots r_i$ is
$\mathcal{J}^r_{ij}(\dots)=k^r_{ij}P(\dots r_j)-k^r_{ji}P(\dots r_i)$,
and analogous for wrong matches (see Fig.~\ref{fig:general}A). The
master equations for the intermediate states can be expressed in a compact
form in terms of these fluxes
\begin{eqnarray}
\label{eq:intermediate}
\dot{P}(\dots r_i)=\sum\limits_{j=0}^{n+1} \mathcal{J}^r_{ij}(\dots)\;\; ,\;\;
\dot{P}(\dots w_i)=\sum\limits_{j=0}^{n+1}\mathcal{J}^w_{ij}(\dots) 
\end{eqnarray}
where the upper dot denotes time derivative. Note that the sum extends
to $j=0$ and $j=n+1$, which with an abuse of notation
correspond to the main states neighboring the network of intermediate
states, $\dots r_0\equiv\dots w_0\equiv\dots\ $, $\dots r_{n+1}\equiv\dots r$ and
$\dots w_{n+1}\equiv\dots w$.  Master equations for main states are 
easily written by distinguishing states ending with a wrong match
from those ending with a right match 
\begin{eqnarray}\label{eq:main}
  \dot{P}(\dots w)&=&\sum\limits_{j=0}^{n+1}\left[ \mathcal{J}^w_{n+1j}(\dots)
    -\mathcal{J}^r_{j0}(\dots w)-\mathcal{J}^w_{j0}(\dots w)\right],\nonumber\\
  \dot{P}(\dots
  r)&=&\sum\limits_{j=0}^{n+1}\left[\mathcal{J}^r_{n+1j}(\dots)
    -\mathcal{J}^r_{j0}(\dots r)-\mathcal{J}^w_{j0}(\dots r)\right]
\end{eqnarray}
where the three sets of fluxes in each equation correspond to
finalized incorporation of the last monomer in the main state, and
attempted incorporation of a right and wrong monomer.  Eqs. (\ref{eq:intermediate}) are similar to those
written for biochemical models, while
Eqs. (\ref{eq:main}) are similar to those used for polymer
growth.

The system of equations (\ref{eq:intermediate}) and (\ref{eq:main})
can be solved at steady state, $\dot{P}=0$, by means of the {\it
  ansatz} that errors are uncorrelated. Given an error $\eta$, to be
determined {\it a posteriori}, we impose that the steady-state
probability of a string of length $N$ with $N^w$ errors is
$P(\dots)\propto\eta^{N^w}(1-\eta)^{N-N^w}$. This implies
\begin{equation}
P(\dots r)=P(\dots)(1-\eta)\;\;{\rm and}\;\; P(\dots w)=P(\dots)\eta\;.
\label{eq:ansatz1}
\end{equation}
For the intermediate states we make the additional {\it ansatz}
\begin{align}
P(\dots r_i)=P(\dots)p^r_i\;\;{\rm and}\;\; P(\dots w_i)=P(\dots)p^w_i\;,
\label{eq:ansatz2}
\end{align}
where $p_i^r$ and $p_i^w$ are the occupancies of the intermediate states
$1\le i\le n$, assumed to be independent of $P(\dots)$.

Substituting Eqs. \ref{eq:ansatz1} and \ref{eq:ansatz2} in
\ref{eq:intermediate} yields a system of $2n$ linear equations, from
which the occupancies can be expressed as functions of the kinetic
rates and the error $\eta$, still to be determined. It is now convenient
to define the occupation fluxes $J_{ij}^r$ as
\begin{equation}
\label{eq:occflux}
J^r_{ij}=\mathcal{N}(k_{ij}^r p_j -k_{ji}^r p_i)\quad,
\end{equation}
where 
$\mathcal{N}=\left[1+\sum_{i=1}^{n}(p^r_i+p_i^w)\right]^{-1}$ is a
normalization constant such that $\sum_{\ldots i} P(\ldots r_i)+P(\ldots w_i)=1$, and thus $\sum_{\dots}P(\ldots)=\mathcal{N}$.
Occupation fluxes are related to the probability fluxes via
$\mathcal{J}^r_{ij}(\dots)=P(\dots) J^r_{ij}/\mathcal{N}$ and
analogously for wrong matches. The speed of right and wrong monomer
incorporations can now be expressed as $v^r =
\sum_iJ^r_{n+1i}=\sum_iJ^r_{i0}$ and
$v^w=\sum_iJ^w_{n+1i}=\sum_iJ^w_{i0}$. Replacing the {\it ansatz}
in Eqs. \ref{eq:main} and using these definitions
results in Eq.~\ref{eq:error}, which can be finally used to determine
the error.

\subsection{Entropy production rate} To calculate the steady-state entropy production rate, we start with the general expression \cite{Schnakenberg1974}
\begin{eqnarray}\label{algo1}
\dot{S}_{\rm tot}=\frac12\sum_{\dots,i,j}\left[\mathcal{J}^r_{ij}(\dots)\log\left(\frac{k^r_{ij}P(\dots
      r_j)}{k^r_{ji}P(\dots r_i)}\right)\right. +\nonumber\\
\left. + \mathcal{J}^w_{ij}(\dots)\log\left(\frac{k^w_{ij}P(\dots w_j)}{k^w_{ji}P(\dots w_i)}\right) \right].
\end{eqnarray}
We now factorize the sum into one over strings (noted $\sum_{\dots}$)
and one over intermediate states (where $\langle ij\rangle$ denotes
links). Using the definition of the occupation fluxes,
Eq. \ref{eq:occflux}, we obtain:
\begin{align}\label{algo2}
\dot{S}_{\rm tot}=\frac{\sum_{\dots}P(\dots)}{\mathcal{N}}\sum_{\langle ij\rangle}&\Bigg[J^r_{ij}\log\left(\frac{k^r_{ij}p^r_j}{k^r_{ji}p^r_i}\right)\nonumber\\&+J^w_{ij}\log\left(\frac{k^w_{ij}{p}^w_j}{k^w_{ji}{p}^w_i}\right)\Bigg].
\end{align}
Since the sum over all states is normalized to one, we have that $\sum_{\dots}P(\dots)=\left[1+\sum_{i=1}^{n}(p^r_i+p_i^w)\right]^{-1}$. Using the definition of $\mathcal{N}$ in previous section, the term outside the brackets is equal to $1$.
Substituting the definition of the rates of Fig. (\ref{fig:rates}) into (\ref{algo2}) yields
\begin{eqnarray}
\label{eq:epr}
\dot{S}_{\rm tot}&=&\sum_{\langle ij\rangle}
(J^r_{ij}+J^w_{ij})\mu_{ij}/T+\sum_{\langle
  ij\rangle}J^r_{ij}\log\left(\frac{p^r_j}{ p^r_i}\right)\nonumber\\
&+&\sum_{\langle ij\rangle}J^w_{ij}\log\left(\frac{p^w_j}{
    p^w_i}\right) +\sum_{\langle ij\rangle}J^r_{ij}\left(\Delta E^r_j-\Delta E^r_i\right)/T\nonumber\\
&+&\sum_{\langle ij\rangle}J^w_{ij}\left(\Delta E^w_j -\Delta E^r_i \right)/T\quad.
\end{eqnarray}
For an isolated network at steady state, all terms but the first one vanish by flux conservation \cite{Schnakenberg1974}. However, in cyclic copying the states $i=0$ and $i=n+1$ receive a finite flux from the rest of the transition network, see Fig. \ref{fig:general}A. Using $\sum_j J_{ij}^r=0$ for $1\le i\le n$, the definitions of $v^r$ and $v^w$, and Eq.~\ref{eq:error}, we obtain
 \begin{eqnarray}\label{algo}
 \dot{S}_{\rm tot}&=&\sum_{\langle ij\rangle}(J^r_{ij}+J^w_{ij})\mu_{ij}/T-\eta v[\log\left(\eta\right) + \Delta E^w/T]\nonumber\\
 &-&(1-\eta)v[\log\left(1-\eta\right) + \Delta E^r/T]
 \end{eqnarray}
Using the definition of equilibrium error and free energy difference per step given in {\em Results}, we arrive at
\begin{equation}
T\dot{S}_{\rm tot}=v[\Delta W - \Delta F_{\rm eq}-T~D(\eta||\eta_{\rm eq})]\quad.
\end{equation}
Defining the entropy production per step as $\Delta S_{\rm
tot}=\dot{S}_{\rm tot}/v$ leads to Eq. \ref{eq:entropyprod}.

Eq. (\ref{eq:ESA}) can be derived following the same procedure, but
considering the contribution to the entropy production coming
  from incorporation of wrong matches, $\dot{S}^w_{\rm
    tot}=\frac12\sum_{\dots,i,j}
  \mathcal{J}^w_{ij}(\dots)\log[k^w_{ij}P(\dots w_j)/(k^w_{ji}P(\dots
  w_i)]$, from which we also define $\Delta S_{\rm tot}^w=\dot{S}^w_{\rm
    tot}/v^w$. Note that $\dot{S}^w_{\rm tot}\ge 0$, since all terms
  of the sum in its definition are non-negative.

\subsection{Thermodynamic bound for proofreading} 

In copying schemes assisted by kinetic proofreading the proofreading
reaction removes incorporated monomers at an average speed {\cred
  $v_{\rm p} = J^w_{n+1\ 0}+ J^r_{n+1\ 0}$, where the subindex ``p''
  denotes quantities that correspond to the proofreading reaction. The
  average proofreading speed can be written as a sum of contributions
  coming from right and wrong monomers $v_{\rm p}=v_{\rm p}^r+v_{\rm
    p}^w<0$. Proofreading is fueled by a chemical driving $\mu_{0\
    n+1}$, which is the same for right and wrong matches (we remind
  that the proofreading reaction is driven backward). By direct
  substitution, one can show that the average work per proofread
  monomer is $\Delta{W}_{\rm p}=\Delta{W}_{\rm p}^w=\Delta{W}_{\rm
    p}^r=\mu_{0\ n+1 }$ } According to our convention, monomer removal
corresponds to $v_{\rm p}<0$.  In an effective proofreading scheme,
errors are removed on average, $v_{\rm p}^w=J^w_{n+1\ 0}<0$. Consider
now the entropy production rate of proofreading wrong monomers,
$\dot{S}^w_{\rm p, tot}=J^w_{n+1\ 0}\log[(p_0^wk_{
  n+1\ 0}^w)/(p^w_{n+1}k_{0\ n+1})]$. As every term of $\dot{S}_{\rm
  tot}$, this quantity satisfies a second-law-like inequality
$\dot{S}^w_{\rm p, tot}\ge0$. By means of this inequality, and using
$v_{\rm p}^w<0$, $p^w_0=1$ and $p^w_{n+1}=\eta$, we obtain the general
proofreading bound of Eq.~\ref{eq:kpbound}.

\subsection{Solution of the proofreading model} To solve the proofreading protocol in Fig.~\ref{fig:kp}A, we start
from Eqs. \ref{eq:intermediate}, which at steady state imply
$J^r_{10}-J^r_{21}=0$ and $J^w_{10}-J^w_{21}=0$. Solving for the
probabilities of the intermediate states yields
$p_1^r=(k^r_{10}+(1-\eta)k^r_{12})/(k^r_{01}+k^r_{21})$ and
$p_1^w=(k^w_{10}+k^w_{12}\eta)/(k^w_{01}+k^w_{21})$. The speed of
incorporation of right and wrong monomers are
$v^w=\mathcal{N}[k^w_{20}+k^w_{21}p_1^w-\eta(k^w_{12}+k^w_{02})]$ and
$v^r=\mathcal{N}[k^r_{20}+k^r_{21}p_1^r-(1-\eta)(k^r_{12}+k^r_{02})]$,
where $\mathcal{N}$ is the previously defined normalization
constant. Substituting these expressions in Eq. \ref{eq:error} yields
\begin{equation}
\label{eq:kp_hop_sol}
\frac{\eta}{1-\eta}=\frac{k^w_{20}+k^w_{21}p_1^w-\eta(k^w_{12}+k^w_{02})}{k^r_{20}+k^r_{21}p_1^r-(1-\eta)(k^r_{12}+k^r_{02})}\;,
\end{equation}
which can be easily solved for the error $\eta$.

To scrutinize the effectiveness of proofreading, we parametrize the rates as in Fig.~\ref{fig:rates}B. Considering the strongly-driven regime $\mu_{21},\mu_{02}\gg1$,  Eq. (\ref{eq:kp_hop_sol}) becomes
\begin{equation}
\label{eq:kp_hop_sol2}
\frac{\eta}{1-\eta}=\frac{\omega_{21}  p_1^w -\eta  \omega_{02}
  e^{(\mu_{02}-\mu_{21}+\Delta E^w)/T} } {\omega_{21}  p_1^r   -(1-\eta)
  \omega_{02} e^{(\mu_{02}-\mu_{21}+\Delta E^r+\delta_{02})/T}}\;.
\end{equation}
From Eq. \ref{eq:kp_hop_sol2}, one can deduce that the error $\eta$ is
a decreasing function of the combination of parameters
$K=(\omega_{02}/\omega_{21}) e^{(\mu_{02}-\mu_{21})/T}$, which tunes
the intensity of proofreading. However, increasing $K$ also increases
the absolute value of the proofreading speed $v_{\rm
  p}=\mathcal{N}[k^r_{20}+k^w_{20}-k^r_{02}(1-\eta)+k^w_{02}\eta]$, so
that $K$ can be increased only up to a point where the net elongation
speed vanishes. Finding the maximum value of $K$ by the condition
$v=0$ and substituting in Eq. (\ref{eq:kp_hop_sol2}) leads to
Eq. \ref{eq:hardware}. In this case,  $\eta_1$ is determined by the
large kinetic barrier $\eta_1\approx e^{-\delta_{10}/T}$, see
e.g. \cite{cady2009open,sartori2013kinetic}.

\begin{acknowledgments}
  We thank L. Granger, R. Ma, L. Peliti, A. Puglisi and three
  anonymous referees for useful comments on a preliminary version of
  the manuscript. SP acknowledges partial support from Spanish
  research ministry through grant FIS2012-37655-C02-01.
\end{acknowledgments}

\bibliography{thermocopy}

\begin{thebibliography}{31}%
\makeatletter
\providecommand \@ifxundefined [1]{%
 \@ifx{#1\undefined}
}%
\providecommand \@ifnum [1]{%
 \ifnum #1\expandafter \@firstoftwo
 \else \expandafter \@secondoftwo
 \fi
}%
\providecommand \@ifx [1]{%
 \ifx #1\expandafter \@firstoftwo
 \else \expandafter \@secondoftwo
 \fi
}%
\providecommand \natexlab [1]{#1}%
\providecommand \enquote  [1]{``#1''}%
\providecommand \bibnamefont  [1]{#1}%
\providecommand \bibfnamefont [1]{#1}%
\providecommand \citenamefont [1]{#1}%
\providecommand \href@noop [0]{\@secondoftwo}%
\providecommand \href [0]{\begingroup \@sanitize@url \@href}%
\providecommand \@href[1]{\@@startlink{#1}\@@href}%
\providecommand \@@href[1]{\endgroup#1\@@endlink}%
\providecommand \@sanitize@url [0]{\catcode `\\12\catcode `\$12\catcode
  `\&12\catcode `\#12\catcode `\^12\catcode `\_12\catcode `\%12\relax}%
\providecommand \@@startlink[1]{}%
\providecommand \@@endlink[0]{}%
\providecommand \url  [0]{\begingroup\@sanitize@url \@url }%
\providecommand \@url [1]{\endgroup\@href {#1}{\urlprefix }}%
\providecommand \urlprefix  [0]{URL }%
\providecommand \Eprint [0]{\href }%
\providecommand \doibase [0]{http://dx.doi.org/}%
\providecommand \selectlanguage [0]{\@gobble}%
\providecommand \bibinfo  [0]{\@secondoftwo}%
\providecommand \bibfield  [0]{\@secondoftwo}%
\providecommand \translation [1]{[#1]}%
\providecommand \BibitemOpen [0]{}%
\providecommand \bibitemStop [0]{}%
\providecommand \bibitemNoStop [0]{.\EOS\space}%
\providecommand \EOS [0]{\spacefactor3000\relax}%
\providecommand \BibitemShut  [1]{\csname bibitem#1\endcsname}%
\let\auto@bib@innerbib\@empty
\bibitem [{\citenamefont {Von~Neumann}(1956)}]{von1956probabilistic}%
  \BibitemOpen
  \bibfield  {author} {\bibinfo {author} {\bibfnamefont {John}\ \bibnamefont
  {Von~Neumann}},\ }\bibfield  {title} {\enquote {\bibinfo {title}
  {Probabilistic logics and the synthesis of reliable organisms from unreliable
  components},}\ }\href@noop {} {\bibfield  {journal} {\bibinfo  {journal}
  {Automata studies}\ }\textbf {\bibinfo {volume} {34}},\ \bibinfo {pages}
  {43--98} (\bibinfo {year} {1956})}\BibitemShut {NoStop}%
\bibitem [{\citenamefont {Johnson}(1993)}]{johnson1993conformational}%
  \BibitemOpen
  \bibfield  {author} {\bibinfo {author} {\bibfnamefont {Kenneth~A}\
  \bibnamefont {Johnson}},\ }\bibfield  {title} {\enquote {\bibinfo {title}
  {Conformational coupling in dna polymerase fidelity},}\ }\href@noop {}
  {\bibfield  {journal} {\bibinfo  {journal} {Annual review of biochemistry}\
  }\textbf {\bibinfo {volume} {62}},\ \bibinfo {pages} {685--713} (\bibinfo
  {year} {1993})}\BibitemShut {NoStop}%
\bibitem [{\citenamefont {Bennett}(1982)}]{Bennett:1982wx}%
  \BibitemOpen
  \bibfield  {author} {\bibinfo {author} {\bibfnamefont {C.H.}\ \bibnamefont
  {Bennett}},\ }\bibfield  {title} {\enquote {\bibinfo {title} {{The
  thermodynamics of computation---a review}},}\ }\href@noop {} {\bibfield
  {journal} {\bibinfo  {journal} {Int. J. Theor. Phys.}\ }\textbf {\bibinfo
  {volume} {21}},\ \bibinfo {pages} {905--940} (\bibinfo {year}
  {1982})}\BibitemShut {NoStop}%
\bibitem [{\citenamefont {Hopfield}(1974)}]{hopfield1974kinetic}%
  \BibitemOpen
  \bibfield  {author} {\bibinfo {author} {\bibfnamefont {John~J}\ \bibnamefont
  {Hopfield}},\ }\bibfield  {title} {\enquote {\bibinfo {title} {{Kinetic
  proofreading: a new mechanism for reducing errors in biosynthetic processes
  requiring high specificity}},}\ }\href@noop {} {\bibfield  {journal}
  {\bibinfo  {journal} {Proceedings of the National Academy of Sciences}\
  }\textbf {\bibinfo {volume} {71}},\ \bibinfo {pages} {4135--4139} (\bibinfo
  {year} {1974})}\BibitemShut {NoStop}%
\bibitem [{\citenamefont {Ninio}(1975)}]{ninio1975kinetic}%
  \BibitemOpen
  \bibfield  {author} {\bibinfo {author} {\bibfnamefont {Jacques}\ \bibnamefont
  {Ninio}},\ }\bibfield  {title} {\enquote {\bibinfo {title} {{Kinetic
  amplification of enzyme discrimination}},}\ }\href@noop {} {\bibfield
  {journal} {\bibinfo  {journal} {Biochimie}\ }\textbf {\bibinfo {volume}
  {57}},\ \bibinfo {pages} {587--595} (\bibinfo {year} {1975})}\BibitemShut
  {NoStop}%
\bibitem [{\citenamefont {Murugan}\ \emph {et~al.}(2012)\citenamefont
  {Murugan}, \citenamefont {Huse},\ and\ \citenamefont
  {Leibler}}]{murugan2012speed}%
  \BibitemOpen
  \bibfield  {author} {\bibinfo {author} {\bibfnamefont {Arvind}\ \bibnamefont
  {Murugan}}, \bibinfo {author} {\bibfnamefont {David~A}\ \bibnamefont {Huse}},
  \ and\ \bibinfo {author} {\bibfnamefont {Stanislas}\ \bibnamefont
  {Leibler}},\ }\bibfield  {title} {\enquote {\bibinfo {title} {Speed,
  dissipation, and error in kinetic proofreading},}\ }\href@noop {} {\bibfield
  {journal} {\bibinfo  {journal} {Proceedings of the National Academy of
  Sciences}\ }\textbf {\bibinfo {volume} {109}},\ \bibinfo {pages}
  {12034--12039} (\bibinfo {year} {2012})}\BibitemShut {NoStop}%
\bibitem [{\citenamefont {Murugan}\ \emph {et~al.}(2014)\citenamefont
  {Murugan}, \citenamefont {Huse},\ and\ \citenamefont
  {Leibler}}]{murugan2014discriminatory}%
  \BibitemOpen
  \bibfield  {author} {\bibinfo {author} {\bibfnamefont {Arvind}\ \bibnamefont
  {Murugan}}, \bibinfo {author} {\bibfnamefont {David~A}\ \bibnamefont {Huse}},
  \ and\ \bibinfo {author} {\bibfnamefont {Stanislas}\ \bibnamefont
  {Leibler}},\ }\bibfield  {title} {\enquote {\bibinfo {title} {Discriminatory
  proofreading regimes in nonequilibrium systems},}\ }\href@noop {} {\bibfield
  {journal} {\bibinfo  {journal} {Physical Review X}\ }\textbf {\bibinfo
  {volume} {4}},\ \bibinfo {pages} {021016} (\bibinfo {year}
  {2014})}\BibitemShut {NoStop}%
\bibitem [{\citenamefont {Bennett}(1979)}]{bennett1979dissipation}%
  \BibitemOpen
  \bibfield  {author} {\bibinfo {author} {\bibfnamefont {Charles~H}\
  \bibnamefont {Bennett}},\ }\bibfield  {title} {\enquote {\bibinfo {title}
  {{Dissipation-error tradeoff in proofreading}},}\ }\href@noop {} {\bibfield
  {journal} {\bibinfo  {journal} {BioSystems}\ }\textbf {\bibinfo {volume}
  {11}},\ \bibinfo {pages} {85--91} (\bibinfo {year} {1979})}\BibitemShut
  {NoStop}%
\bibitem [{\citenamefont {Andrieux}\ and\ \citenamefont
  {Gaspard}(2008)}]{andrieux2008nonequilibrium}%
  \BibitemOpen
  \bibfield  {author} {\bibinfo {author} {\bibfnamefont {David}\ \bibnamefont
  {Andrieux}}\ and\ \bibinfo {author} {\bibfnamefont {Pierre}\ \bibnamefont
  {Gaspard}},\ }\bibfield  {title} {\enquote {\bibinfo {title} {Nonequilibrium
  generation of information in copolymerization processes},}\ }\href@noop {}
  {\bibfield  {journal} {\bibinfo  {journal} {Proceedings of the National
  Academy of Sciences}\ }\textbf {\bibinfo {volume} {105}},\ \bibinfo {pages}
  {9516--9521} (\bibinfo {year} {2008})}\BibitemShut {NoStop}%
\bibitem [{\citenamefont {Cady}\ and\ \citenamefont
  {Qian}(2009)}]{cady2009open}%
  \BibitemOpen
  \bibfield  {author} {\bibinfo {author} {\bibfnamefont {Field}\ \bibnamefont
  {Cady}}\ and\ \bibinfo {author} {\bibfnamefont {Hong}\ \bibnamefont {Qian}},\
  }\bibfield  {title} {\enquote {\bibinfo {title} {Open-system thermodynamic
  analysis of dna polymerase fidelity},}\ }\href@noop {} {\bibfield  {journal}
  {\bibinfo  {journal} {Physical biology}\ }\textbf {\bibinfo {volume} {6}},\
  \bibinfo {pages} {036011} (\bibinfo {year} {2009})}\BibitemShut {NoStop}%
\bibitem [{\citenamefont {Andrieux}\ and\ \citenamefont
  {Gaspard}(2009)}]{andrieux2009molecular}%
  \BibitemOpen
  \bibfield  {author} {\bibinfo {author} {\bibfnamefont {David}\ \bibnamefont
  {Andrieux}}\ and\ \bibinfo {author} {\bibfnamefont {Pierre}\ \bibnamefont
  {Gaspard}},\ }\bibfield  {title} {\enquote {\bibinfo {title} {Molecular
  information processing in nonequilibrium copolymerizations},}\ }\href@noop {}
  {\bibfield  {journal} {\bibinfo  {journal} {The Journal of chemical physics}\
  }\textbf {\bibinfo {volume} {130}},\ \bibinfo {pages} {014901} (\bibinfo
  {year} {2009})}\BibitemShut {NoStop}%
\bibitem [{\citenamefont {Esposito}\ \emph {et~al.}(2010)\citenamefont
  {Esposito}, \citenamefont {Lindenberg},\ and\ \citenamefont {Van~den
  Broeck}}]{esposito2010extracting}%
  \BibitemOpen
  \bibfield  {author} {\bibinfo {author} {\bibfnamefont {Massimiliano}\
  \bibnamefont {Esposito}}, \bibinfo {author} {\bibfnamefont {Katja}\
  \bibnamefont {Lindenberg}}, \ and\ \bibinfo {author} {\bibfnamefont
  {Christian}\ \bibnamefont {Van~den Broeck}},\ }\bibfield  {title} {\enquote
  {\bibinfo {title} {{Extracting chemical energy by growing disorder:
  efficiency at maximum power}},}\ }\href@noop {} {\bibfield  {journal}
  {\bibinfo  {journal} {Journal of Statistical Mechanics: Theory and
  Experiment}\ }\textbf {\bibinfo {volume} {2010}},\ \bibinfo {pages} {P01008}
  (\bibinfo {year} {2010})}\BibitemShut {NoStop}%
\bibitem [{\citenamefont {Sartori}\ and\ \citenamefont
  {Pigolotti}(2013)}]{sartori2013kinetic}%
  \BibitemOpen
  \bibfield  {author} {\bibinfo {author} {\bibfnamefont {Pablo}\ \bibnamefont
  {Sartori}}\ and\ \bibinfo {author} {\bibfnamefont {Simone}\ \bibnamefont
  {Pigolotti}},\ }\bibfield  {title} {\enquote {\bibinfo {title} {{Kinetic
  versus energetic discrimination in biological copying}},}\ }\href@noop {}
  {\bibfield  {journal} {\bibinfo  {journal} {Physical review letters}\
  }\textbf {\bibinfo {volume} {110}},\ \bibinfo {pages} {188101} (\bibinfo
  {year} {2013})}\BibitemShut {NoStop}%
\bibitem [{\citenamefont {Andrieux}\ and\ \citenamefont
  {Gaspard}(2013)}]{andrieux2013information}%
  \BibitemOpen
  \bibfield  {author} {\bibinfo {author} {\bibfnamefont {David}\ \bibnamefont
  {Andrieux}}\ and\ \bibinfo {author} {\bibfnamefont {Pierre}\ \bibnamefont
  {Gaspard}},\ }\bibfield  {title} {\enquote {\bibinfo {title} {Information
  erasure in copolymers},}\ }\href@noop {} {\bibfield  {journal} {\bibinfo
  {journal} {EPL (Europhysics Letters)}\ }\textbf {\bibinfo {volume} {103}},\
  \bibinfo {pages} {30004} (\bibinfo {year} {2013})}\BibitemShut {NoStop}%
\bibitem [{\citenamefont {Gaspard}\ and\ \citenamefont
  {Andrieux}(2014)}]{gaspard2014kinetics}%
  \BibitemOpen
  \bibfield  {author} {\bibinfo {author} {\bibfnamefont {Pierre}\ \bibnamefont
  {Gaspard}}\ and\ \bibinfo {author} {\bibfnamefont {David}\ \bibnamefont
  {Andrieux}},\ }\bibfield  {title} {\enquote {\bibinfo {title} {{Kinetics and
  thermodynamics of first-order Markov chain copolymerization}},}\ }\href@noop
  {} {\bibfield  {journal} {\bibinfo  {journal} {The Journal of chemical
  physics}\ }\textbf {\bibinfo {volume} {141}},\ \bibinfo {pages} {044908}
  (\bibinfo {year} {2014})}\BibitemShut {NoStop}%
\bibitem [{\citenamefont {Sharma}\ and\ \citenamefont
  {Chowdhury}(2012)}]{sharma2012template}%
  \BibitemOpen
  \bibfield  {author} {\bibinfo {author} {\bibfnamefont {Ajeet~K}\ \bibnamefont
  {Sharma}}\ and\ \bibinfo {author} {\bibfnamefont {Debashish}\ \bibnamefont
  {Chowdhury}},\ }\bibfield  {title} {\enquote {\bibinfo {title}
  {Template-directed biopolymerization: tape-copying turing machines},}\
  }\href@noop {} {\bibfield  {journal} {\bibinfo  {journal} {Biophysical
  Reviews and Letters}\ }\textbf {\bibinfo {volume} {7}},\ \bibinfo {pages}
  {135--175} (\bibinfo {year} {2012})}\BibitemShut {NoStop}%
\bibitem [{\citenamefont {Johansson}\ \emph {et~al.}(2008)\citenamefont
  {Johansson}, \citenamefont {Lovmar},\ and\ \citenamefont
  {Ehrenberg}}]{johansson2008rate}%
  \BibitemOpen
  \bibfield  {author} {\bibinfo {author} {\bibfnamefont {Magnus}\ \bibnamefont
  {Johansson}}, \bibinfo {author} {\bibfnamefont {Martin}\ \bibnamefont
  {Lovmar}}, \ and\ \bibinfo {author} {\bibfnamefont {M{\aa}ns}\ \bibnamefont
  {Ehrenberg}},\ }\bibfield  {title} {\enquote {\bibinfo {title} {Rate and
  accuracy of bacterial protein synthesis revisited},}\ }\href@noop {}
  {\bibfield  {journal} {\bibinfo  {journal} {Current opinion in microbiology}\
  }\textbf {\bibinfo {volume} {11}},\ \bibinfo {pages} {141--147} (\bibinfo
  {year} {2008})}\BibitemShut {NoStop}%
\bibitem [{\citenamefont {Pape}\ \emph {et~al.}(1999)\citenamefont {Pape},
  \citenamefont {Wintermeyer},\ and\ \citenamefont
  {Rodnina}}]{pape1999induced}%
  \BibitemOpen
  \bibfield  {author} {\bibinfo {author} {\bibfnamefont {Tillmann}\
  \bibnamefont {Pape}}, \bibinfo {author} {\bibfnamefont {Wolfgang}\
  \bibnamefont {Wintermeyer}}, \ and\ \bibinfo {author} {\bibfnamefont
  {Marina}\ \bibnamefont {Rodnina}},\ }\bibfield  {title} {\enquote {\bibinfo
  {title} {Induced fit in initial selection and proofreading of aminoacyl-trna
  on the ribosome},}\ }\href@noop {} {\bibfield  {journal} {\bibinfo  {journal}
  {The EMBO Journal}\ }\textbf {\bibinfo {volume} {18}},\ \bibinfo {pages}
  {3800--3807} (\bibinfo {year} {1999})}\BibitemShut {NoStop}%
\bibitem [{\citenamefont {Zaher}\ and\ \citenamefont
  {Green}(2009)}]{zaher2009fidelity}%
  \BibitemOpen
  \bibfield  {author} {\bibinfo {author} {\bibfnamefont {Hani~S}\ \bibnamefont
  {Zaher}}\ and\ \bibinfo {author} {\bibfnamefont {Rachel}\ \bibnamefont
  {Green}},\ }\bibfield  {title} {\enquote {\bibinfo {title} {Fidelity at the
  molecular level: lessons from protein synthesis},}\ }\href@noop {} {\bibfield
   {journal} {\bibinfo  {journal} {Cell}\ }\textbf {\bibinfo {volume} {136}},\
  \bibinfo {pages} {746--762} (\bibinfo {year} {2009})}\BibitemShut {NoStop}%
\bibitem [{\citenamefont {Mandal}\ and\ \citenamefont
  {Jarzynski}(2012)}]{jarz}%
  \BibitemOpen
  \bibfield  {author} {\bibinfo {author} {\bibfnamefont {Dibyendu}\
  \bibnamefont {Mandal}}\ and\ \bibinfo {author} {\bibfnamefont {Christopher}\
  \bibnamefont {Jarzynski}},\ }\bibfield  {title} {\enquote {\bibinfo {title}
  {Work and information processing in a solvable model of maxwell’s demon},}\
  }\href@noop {} {\bibfield  {journal} {\bibinfo  {journal} {Proceedings of the
  National Academy of Sciences}\ }\textbf {\bibinfo {volume} {109}},\ \bibinfo
  {pages} {11641--11645} (\bibinfo {year} {2012})}\BibitemShut {NoStop}%
\bibitem [{\citenamefont {Esposito}\ and\ \citenamefont {Van~den
  Broeck}(2011)}]{esposito_landauer}%
  \BibitemOpen
  \bibfield  {author} {\bibinfo {author} {\bibfnamefont {Massimiliano}\
  \bibnamefont {Esposito}}\ and\ \bibinfo {author} {\bibfnamefont {Christian}\
  \bibnamefont {Van~den Broeck}},\ }\bibfield  {title} {\enquote {\bibinfo
  {title} {Second law and landauer principle far from equilibrium},}\
  }\href@noop {} {\bibfield  {journal} {\bibinfo  {journal} {EPL (Europhysics
  Letters)}\ }\textbf {\bibinfo {volume} {95}},\ \bibinfo {pages} {40004}
  (\bibinfo {year} {2011})}\BibitemShut {NoStop}%
\bibitem [{\citenamefont {Ehrenberg}\ and\ \citenamefont
  {Blomberg}(1980)}]{ehrenberg_proof}%
  \BibitemOpen
  \bibfield  {author} {\bibinfo {author} {\bibfnamefont {M}~\bibnamefont
  {Ehrenberg}}\ and\ \bibinfo {author} {\bibfnamefont {C}~\bibnamefont
  {Blomberg}},\ }\bibfield  {title} {\enquote {\bibinfo {title} {Thermodynamic
  constraints on kinetic proofreading in biosynthetic pathways},}\ }\href@noop
  {} {\bibfield  {journal} {\bibinfo  {journal} {Biophysical journal}\ }\textbf
  {\bibinfo {volume} {31}},\ \bibinfo {pages} {333--358} (\bibinfo {year}
  {1980})}\BibitemShut {NoStop}%
\bibitem [{\citenamefont {Qian}(2006)}]{qian_noise}%
  \BibitemOpen
  \bibfield  {author} {\bibinfo {author} {\bibfnamefont {Hong}\ \bibnamefont
  {Qian}},\ }\bibfield  {title} {\enquote {\bibinfo {title} {Reducing intrinsic
  biochemical noise in cells and its thermodynamic limit},}\ }\href@noop {}
  {\bibfield  {journal} {\bibinfo  {journal} {Journal of molecular biology}\
  }\textbf {\bibinfo {volume} {362}},\ \bibinfo {pages} {387--392} (\bibinfo
  {year} {2006})}\BibitemShut {NoStop}%
\bibitem [{\citenamefont {Galburt}\ \emph {et~al.}(2007)\citenamefont
  {Galburt}, \citenamefont {Grill}, \citenamefont {Wiedmann}, \citenamefont
  {Lubkowska}, \citenamefont {Choy}, \citenamefont {Nogales}, \citenamefont
  {Kashlev},\ and\ \citenamefont {Bustamante}}]{galburt2007backtracking}%
  \BibitemOpen
  \bibfield  {author} {\bibinfo {author} {\bibfnamefont {Eric~A}\ \bibnamefont
  {Galburt}}, \bibinfo {author} {\bibfnamefont {Stephan~W}\ \bibnamefont
  {Grill}}, \bibinfo {author} {\bibfnamefont {Anna}\ \bibnamefont {Wiedmann}},
  \bibinfo {author} {\bibfnamefont {Lucyna}\ \bibnamefont {Lubkowska}},
  \bibinfo {author} {\bibfnamefont {Jason}\ \bibnamefont {Choy}}, \bibinfo
  {author} {\bibfnamefont {Eva}\ \bibnamefont {Nogales}}, \bibinfo {author}
  {\bibfnamefont {Mikhail}\ \bibnamefont {Kashlev}}, \ and\ \bibinfo {author}
  {\bibfnamefont {Carlos}\ \bibnamefont {Bustamante}},\ }\bibfield  {title}
  {\enquote {\bibinfo {title} {{Backtracking determines the force sensitivity
  of RNAP II in a factor-dependent manner}},}\ }\href@noop {} {\bibfield
  {journal} {\bibinfo  {journal} {Nature}\ }\textbf {\bibinfo {volume} {446}},\
  \bibinfo {pages} {820--823} (\bibinfo {year} {2007})}\BibitemShut {NoStop}%
\bibitem [{\citenamefont {Depken}\ \emph {et~al.}(2013)\citenamefont {Depken},
  \citenamefont {Parrondo},\ and\ \citenamefont {Grill}}]{depken}%
  \BibitemOpen
  \bibfield  {author} {\bibinfo {author} {\bibfnamefont {Martin}\ \bibnamefont
  {Depken}}, \bibinfo {author} {\bibfnamefont {Juan~MR}\ \bibnamefont
  {Parrondo}}, \ and\ \bibinfo {author} {\bibfnamefont {Stephan~W}\
  \bibnamefont {Grill}},\ }\bibfield  {title} {\enquote {\bibinfo {title}
  {Intermittent transcription dynamics for the rapid production of long
  transcripts of high fidelity},}\ }\href@noop {} {\bibfield  {journal}
  {\bibinfo  {journal} {Cell reports}\ }\textbf {\bibinfo {volume} {5}},\
  \bibinfo {pages} {521--530} (\bibinfo {year} {2013})}\BibitemShut {NoStop}%
\bibitem [{\citenamefont {Mellenius}\ and\ \citenamefont
  {Ehrenberg}(2015)}]{mellenius2015dna}%
  \BibitemOpen
  \bibfield  {author} {\bibinfo {author} {\bibfnamefont {Harriet}\ \bibnamefont
  {Mellenius}}\ and\ \bibinfo {author} {\bibfnamefont {M{\aa}ns}\ \bibnamefont
  {Ehrenberg}},\ }\bibfield  {title} {\enquote {\bibinfo {title} {{DNA Template
  Dependent Accuracy Variation of Nucleotide Selection in Transcription}},}\
  }\href@noop {} {\bibfield  {journal} {\bibinfo  {journal} {PloS one}\
  }\textbf {\bibinfo {volume} {10}},\ \bibinfo {pages} {e0119588} (\bibinfo
  {year} {2015})}\BibitemShut {NoStop}%
\bibitem [{\citenamefont {Lan}\ \emph {et~al.}(2012)\citenamefont {Lan},
  \citenamefont {Sartori}, \citenamefont {Neumann}, \citenamefont {Sourjik},\
  and\ \citenamefont {Tu}}]{lan2012energy}%
  \BibitemOpen
  \bibfield  {author} {\bibinfo {author} {\bibfnamefont {Ganhui}\ \bibnamefont
  {Lan}}, \bibinfo {author} {\bibfnamefont {Pablo}\ \bibnamefont {Sartori}},
  \bibinfo {author} {\bibfnamefont {Silke}\ \bibnamefont {Neumann}}, \bibinfo
  {author} {\bibfnamefont {Victor}\ \bibnamefont {Sourjik}}, \ and\ \bibinfo
  {author} {\bibfnamefont {Yuhai}\ \bibnamefont {Tu}},\ }\bibfield  {title}
  {\enquote {\bibinfo {title} {The energy-speed-accuracy trade-off in sensory
  adaptation},}\ }\href@noop {} {\bibfield  {journal} {\bibinfo  {journal}
  {Nature physics}\ }\textbf {\bibinfo {volume} {8}},\ \bibinfo {pages}
  {422--428} (\bibinfo {year} {2012})}\BibitemShut {NoStop}%
\bibitem [{\citenamefont {Mora}(2015)}]{morathierry}%
  \BibitemOpen
  \bibfield  {author} {\bibinfo {author} {\bibfnamefont {Thierry}\ \bibnamefont
  {Mora}},\ }\bibfield  {title} {\enquote {\bibinfo {title} {Physical limit to
  concentration sensing amid spurious ligands},}\ }\href {\doibase
  10.1103/PhysRevLett.115.038102} {\bibfield  {journal} {\bibinfo  {journal}
  {Phys. Rev. Lett.}\ }\textbf {\bibinfo {volume} {115}},\ \bibinfo {pages}
  {038102} (\bibinfo {year} {2015})}\BibitemShut {NoStop}%
\bibitem [{\citenamefont {Lalanne}\ and\ \citenamefont
  {Fran{\c{c}}ois}(2015)}]{lalanne2015chemodetection}%
  \BibitemOpen
  \bibfield  {author} {\bibinfo {author} {\bibfnamefont {Jean-Beno{\^\i}t}\
  \bibnamefont {Lalanne}}\ and\ \bibinfo {author} {\bibfnamefont {Paul}\
  \bibnamefont {Fran{\c{c}}ois}},\ }\bibfield  {title} {\enquote {\bibinfo
  {title} {Chemodetection in fluctuating environments: Receptor coupling,
  buffering, and antagonism},}\ }\href@noop {} {\bibfield  {journal} {\bibinfo
  {journal} {Proceedings of the National Academy of Sciences}\ }\textbf
  {\bibinfo {volume} {112}},\ \bibinfo {pages} {1898--1903} (\bibinfo {year}
  {2015})}\BibitemShut {NoStop}%
\bibitem [{\citenamefont {Parrondo}\ \emph {et~al.}(2015)\citenamefont
  {Parrondo}, \citenamefont {Horowitz},\ and\ \citenamefont
  {Sagawa}}]{Parrondo2015}%
  \BibitemOpen
  \bibfield  {author} {\bibinfo {author} {\bibfnamefont {Juan~MR}\ \bibnamefont
  {Parrondo}}, \bibinfo {author} {\bibfnamefont {Jordan~M}\ \bibnamefont
  {Horowitz}}, \ and\ \bibinfo {author} {\bibfnamefont {Takahiro}\ \bibnamefont
  {Sagawa}},\ }\bibfield  {title} {\enquote {\bibinfo {title} {Thermodynamics
  of information},}\ }\href@noop {} {\bibfield  {journal} {\bibinfo  {journal}
  {Nature Physics}\ }\textbf {\bibinfo {volume} {11}},\ \bibinfo {pages}
  {131--139} (\bibinfo {year} {2015})}\BibitemShut {NoStop}%
\bibitem [{\citenamefont {Schnakenberg}(1976)}]{Schnakenberg1974}%
  \BibitemOpen
  \bibfield  {author} {\bibinfo {author} {\bibfnamefont {J}~\bibnamefont
  {Schnakenberg}},\ }\bibfield  {title} {\enquote {\bibinfo {title} {Network
  theory of microscopic and macroscopic behavior of master equation systems},}\
  }\href@noop {} {\bibfield  {journal} {\bibinfo  {journal} {Reviews of Modern
  physics}\ }\textbf {\bibinfo {volume} {48}},\ \bibinfo {pages} {571}
  (\bibinfo {year} {1976})}\BibitemShut {NoStop}%
\end{thebibliography}%

\end{document}